\documentclass{svjour2}                    
\smartqed  
\usepackage{graphicx,color,xspace,hyperref}
\usepackage{amsmath,amssymb}
\usepackage[numbers,sort&compress,square]{natbib}

\begin{document}

\title{Condensed matter lessons about the origin of time}


\author{Gil Jannes}


\institute{Gil Jannes \at
              Modelling \& Numerical Simulation Group, Universidad Carlos III de Madrid, Avda. de la Universidad 30, 28911 Legan\'{e}s, Spain \\
              \email{gil.jannes@uc3m.es}           
}

\date{Received: date / Accepted: date}

\maketitle

\begin{abstract}
It is widely hoped that quantum gravity will shed light on the question of the origin of time in physics. The currently dominant approaches to a candidate quantum theory of gravity have naturally evolved from general relativity, on the one hand, and from particle physics, on the other hand. A third important branch of 20th century `fundamental' physics, condensed-matter physics, also offers an interesting perspective on quantum gravity, and thereby on the problem of time. The bottomline might sound disappointing: to understand the origin of time, much more experimental input is needed than what is available today. Moreover it is far from obvious that we will ever find out the true origin of physical time, even if we become able to directly probe physics at the Planck scale. But we might learn some interesting lessons about time and the structure of our universe in the process. A first lesson is that there are probably several characteristic scales associated with ``quantum gravity'' effects, rather than the single Planck scale usually considered. These can differ by several orders of magnitude, and thereby conspire to hide certain effects expected from quantum gravity, rendering them undetectable even with Planck-scale experiments. A more tentative conclusion is that the hierarchy between general relativity, special relativity and Newtonian physics, usually taken for granted, might have to be interpreted with caution.
\end{abstract}

\section{Introduction}
\label{S:intro}
The pursuit of a quantum theory of gravity is often presented as a quest for the unification of quantum mechanics and general relativity. The approaches to quantum gravity that are currently most popular, string theory and canonical (loop or spin foam) quantum gravity, reflect those two starting points: string theory is a natural extension of particle physics, with its roots in quantum mechanics and quantum field theory, whereas canonical quantum gravity is an attempt to formulate a quantum theory of gravity which is diffeomorphism invariant by construction. When looking back at the revolution brought about by quantum mechanics, however, particle physics represents only part of the package: it has also been crucial for the development of condensed-matter physics. 

An increasingly popular approach to quantum gravity rests on ideas related to low-temperature condensed-matter physics. The basic idea is
that gravity (and maybe electromagnetism and the other gauge fields) might be an ``emergent phenomenon'',
in the sense of representing a collective macroscopic behaviour resulting from a very different microscopic physics, leading to an effective low-energy description somehow similar to the emergence of collective excitations in condensed matter systems. A basic observation in this respect is the following. The fundamental ``quantum gravity'' theory is generally assumed to have the Planck level as its characteristic scale. Expressed as a temperature: $T_{\rm Pl}\sim 10^{32}$ K. On the other hand, most of the observable universe has temperatures of the order of the cosmic background radiation: $T_{\rm CMB}\sim 2.7K$. Even the interior of a star such as the sun is more than 20 orders of magnitude colder than the Planck temperature, while the LHC operated at roughly 16 orders of magnitude below the Planck scale to detect the Higgs boson. So the degrees of freedom of the quantum vacuum of our universe are effectively frozen out, just like in low-temperature condensed matter systems in a laboratory. The physics that we observe might then well be due to the collective excitations that result from the---comparatively tiny---thermal (or other) perturbations of this vacuum. 

Another observation with respect to emergent spacetimes is the following. Even in relatively simple, weakly interacting condensed matter systems, such as Bose-Einstein condensates in atomic gases, the kinematics of the low-energy excitations or phonons can be described by a relativistic field theory, in which the
curved background spacetime is provided by the collective behaviour of
the condensed part of the constituent atoms. In more
complicated fermionic systems, in particular $^3$He-A,
gravitational and gauge fields emerge as the low-energy bosonic degrees of
freedom together with fermionic quasi-matter in a similar way. All
these emergent components share surprisingly many characteristics with
their counterparts in Einstein gravity and the
standard model of particles. 

In what follows, I will try to summarize some of the basic ideas behind this condensed matter approach to gravity, based mainly on the work of G. Volovik (the seminal reference is the book~\cite{Volovik:2003fe}), and discuss some of the possible lessons that it offers us with respect to the origin of time in our physical universe.

\section{Condensed matter and emergent spacetimes}
\label{S:basics}
An intriguing theorem of mathematical physics~\cite{Unruh:1980cg,Visser:1997ux} shows that the
equation of motion of acoustic perturbations in a perfect (irrotational, inviscid and barotropic) fluid can be written as a d'Alembertian equation in curved spacetime: 
\begin{eqnarray}\label{alembert}%
\square \phi\equiv\frac{1}{\sqrt{-g}} \partial_\mu \sqrt{-g} g^{\mu\nu} \partial_\nu \phi =0~.
\end{eqnarray}%
This is essentially the curved-spacetime version of the usual wave equation in flat Minkowski spacetime:
\begin{eqnarray}\label{wave_equation}%
(-\partial_t^2+c^2\partial_x^2+c^2\partial_y^2+c^2\partial_z^2)\phi=0~.
\end{eqnarray}%
So these acoustic perturbations or phonons in perfect fluids travel along the null geodesics of the effective metric $g_{\mu\nu}$, with $g$ its determinant. In other words, they behave as if they were moving in a relativistic spacetime, whose metric $g_{\mu\nu}$ is determined by the bulk of the fluid, i.e., by the collective behaviour of its constituent atoms. In particular, their experience of time is relativistic in exactly the same sense as the one we know from special and general relativity, including all the observer-dependence of simultaneity and associated paradoxes. This is the more curious because the background system in which this relativistic spacetime emerges can simply be described in Newtonian terms: it consists of a fluid in a laboratory, where all velocities are extremely low and thus relativistic corrections irrelevant, and where there is therefore clearly a preferred time, (imperfectly) indicated by the clocks on the wall of the laboratory.

An interesting sidenote is that, while the effective or acoustic metric 
\begin{eqnarray}\label{metric}
g_{\mu \nu}=\frac{\rho}{c} 
\begin{pmatrix}
v^2-c^2 && -\vec{v}^\text{T}\\
-\vec{v} && \mathbf 1
\end{pmatrix}
\end{eqnarray}
with $\mathbf v$ the velocity vector of the background fluid and $c$ its speed of sound (which is related to the fluid density $\rho$), does not reproduce all possible general relativistic metrics, it does provide the possibility to simulate black holes. Indeed, if one has a normal or subsonic region where $v<c$, and a supersonic region where $v>c$, then the border where $v=c$ (and hence the $g_{tt}$ component of the metric changes sign) defines an acoustic horizon in exactly the same sense as event horizons are defined in general relativistic black holes. 

Building on these observations, the idea developed to study certain aspects of general relativity
and quantum field theory by analogy with such perfect fluid systems~\cite{Barcelo:2005fc}. To take
maximal advantage of the analogy, the  microscopic physics of the fluid system
should be well understood, theoretically and experimentally, even in regimes
where the relativistic description breaks down. Then, full calculations
based on firmly verified and controlled physics are (at least in principle)
possible, even beyond the relativistic regime. Additionally, laboratory
experiments become
feasible that could shed light on issues of high-energy physics. 
The paradigmatic example is that of Bose-Einstein condensates (BECs).
BECs fulfill all the listed conditions, acoustic black holes have been realized in BECs~\cite{Lahav:2009wx}, and (phononic) Hawking radiation was indeed recently detected in a BEC black hole analogue~\cite{Steinhauer:2014dra}.

The simple model considered up to now is certainly not a full-fledged model for quantum gravity. A first reason is the fact mentioned earlier that the metric~\eqref{alembert} does not reproduce all possible general relativistic metrics. However, this shortcoming could partly be argued away since it is far from obvious that all mathematical solutions of general relativity represent physically realistic spacetimes. A second and more important problem regards the dynamics of the system. If one wants to extend the analogy with general relativity beyond the kinematical aspects of quantum field
theory in a curved spacetime, there would have to be some way of emulating the
Einstein field equations. However, in a real perfect fluid such as a BEC, the inherent hydrodynamics of the system dominates and completely obfuscates any possible Einsteinian gravitational dynamics based, e.g., on Sakharov's idea of induced gravity~\cite{Sakharov:1967pk}. This problem is crucial for the general idea of obtaining a satisfactory model of emergent gravity. It is also directly related to the question of time, and so I will treat it separately in Sec.~\ref{S:dynamics} below. The bottom line is that, although interesting progress is being made on the problem of studying the ``analogue gravitational dynamics'' of condensed-matter-like systems (see e.g.~\cite{Girelli:2008gc,Girelli:2008qp}, the models studied so far are clearly not sufficiently complex to reproduce the Einstein dynamics of general relativity. However, even if at the moment not reproducing the Einstein equations, the idea that gravity might emerge from an underlying microscopic ``condensed-matter-like'' quantum system has at least two additional trumps to play.

First, in more complicated fermionic systems with a Fermi-point (or Weyl) topology\footnote{Weyl topologies include Fermi points in 3+1 dimensions, and Dirac points in 2+1 dimensions.}, and in
particular $^3$He-A~\cite{Volovik:2003fe}, fermionic quasi-matter emerges at low energy together with effective bosonic gauge and gravitational fields from the quantum vacuum. The construction here is slightly
more involved than in the simple case of BECs, but its essence can be understood as follows. 
The Fermi point is the point in momentum space where the quasi-particle energy is zero. Spatial and temporal perturbations do not destroy the Fermi point, because of its topological stability. They only lead to a general deformation of the energy spectrum near the Fermi point, determined by
\begin{eqnarray}
 g^{\mu\nu}(p_\mu-p_\mu^{(0)})(p_\nu-p_\nu^{(0)})=0~,
\end{eqnarray}
where $g^{\mu\nu}=\eta^{\lambda\sigma}e^\mu_\sigma e^\nu_\sigma$, with $e^\mu_\nu$ the tetrad or vierbein field, and $\eta^{\lambda\sigma}$ the Minkowski metric. So the dynamical change of slope in the energy spectrum near the Fermi point simulates an effective gravitational field $g^{\mu\nu}$ expressed in terms of the tetrad field $e^\mu_\nu$. Note that the effective gravitational field arises as a consequence of a perturbation of the quantum vacuum, and that this leads to a Lorentzian metric $g_{\mu\nu}$, again: even if the underlying system is not Lorentz invariant, and can actually be thought of as Newtonian. 
The quasi-particles move along the geodesics of the effective metric $g_{\mu\nu}$. Moreover, the quasi-particles and gauge fields that emerge from such systems with Fermi-point topology show striking similarities with the ones known from the standard model of particles, including chiral or Weyl fermions and effective quantum electrodynamics: an effective electromagnetic field emerges which reflects changes in the {\it position} of the Fermi point as a consequence of a perturbation of the quantum vacuum, in a similar way 
to how the gravitational field accounts for a change in the {\it slope} of the energy spectrum near the Fermi point~\cite{Volovik:2003fe} (see also~\cite{Barcelo:2014yna}). Thus Volovik has suggested that the condensed-matter
analogy might not be limited to the gravitational sector, but that by carefully
studying the topological properties of quantum vacua, this might also provide a
hint for a ``theory of everything'' that gives a unified description of gravity and
matter, both emerging in the same process.

Second, apart from the issue of unifying quantum mechanics and the general theory of relativity, there is arguably at least one empirical motivation for a quantum theory of gravity: the accelerated expansion of the universe, which seems to imply some form of repulsive ``dark energy''~\cite{Padmanabhan:2008if}. The first intuition from quantum field theory to explain this mysterious repulsive force was that dark energy is simply
the energy of the quantum vacuum, which makes its entry in the Einstein field equations in the
guise of the cosmological constant~\cite{Weinberg:1988cp}. Infamously, the experimentally obtained value of the cosmological constant turned out to disagree with theoretical estimates of the 
quantum vacuum energy by more than a hundred orders of magnitude, and so this
discrepancy seems to constitute an unsurmountable barrier for such an approach. However,
if one takes the condensed matter analogy seriously, then this intuition might
prove to be right after all~\cite{Volovik:2006bh}. Indeed, the value of the quantum vacuum energy in a
condensed matter system in equilibrium is regulated by
macroscopic thermodynamic principles. The vacuum energy density $\epsilon_{\rm vac}=E_{\rm vac}/V$ of a quantum many-body system relevant for the cosmological constant problem is obtained from the expectation value $E_{\rm vac}=<{\cal H}-\mu{\cal N}>_{\rm vac}$, with ${\cal H}$ the many-body Hamiltonian, $\mu$ the chemical potential and ${\cal N}$ the number operator. The equation of state relating the energy density and the pressure of the vacuum of any quantum many-body system is then simply $\epsilon_{\rm vac}=-p_{\rm vac}$, regardless of whether the vacuum is Lorentz invariant or not. Although Volovik's argument should be well-known by now, it is probably worth summarizing it.

For a Lorentz invariant vacuum, $\rho_{\rm vac}=-P_{\rm vac}$ is the only possible equation of state as a perfect fluid, and so one can immediately see from the thermodynamic Gibbs-Duhem relation 
\begin{equation}\label{gibbs-duhem}
 P=-\epsilon + Ts +\sum_i  \mu_i q_i~~,~~ q_i=\frac{Q_i}{V}
\end{equation}
(with $s$ the specific entropy and the temperature $T=0$ in the vacuum) that the relevant thermodynamic quantity, which plays the role of the vacuum energy is the analog of grand-canonical energy $\rho_{\rm vac}=\epsilon(q_i)  - \sum_i  q_i d\epsilon/dq_i$. Any conserved quantity $Q_i$,
which characterizes the quantum vacuum,  should be
explicitly taken into account together with its corresponding Lagrange
multiplier $\mu_i$. And indeed it is demonstrated that the quantity which enters the cosmological term in Einstein equations is the density of the grand-canonical energy, $\rho_{\rm vac}$, rather than the energy density $\epsilon$. 

By reversing the above argument, one sees that the vacuum equation of state $\rho_{\rm vac}=-P_{\rm
vac}$ is more generally valid~\cite{Volovik:2006bh,Jannes:2011em}. The energy of the vacuum of quantum fields emerging
in a many body condensed matter system is the grand canonical
energy $\rho_{\rm vac}=\epsilon(n)  - nd\epsilon/dn$, where the particle
density $N=nV$ is a conserved quantity and the corresponding Lagrange
multiplier is the chemical potential $\mu=d\epsilon/dn$. The use of the
grand canonical energy here corresponds to the fact that the  many-body
Hamiltonian in second quantization is $\hat H_\text{QFT}=\hat H - \mu \hat N$,
where $\hat H$ is obtained from the Schr\"odinger many-body Hamiltonian and
$\hat N$ is the number operator. The cosmological equation of state  $w= P/\rho$ for the vacuum energy is then again $w=-1$  due to the Gibbs-Duhem relation, regardless of the Lorentz invariance or not of the vacuum.

Liquid-like systems can be in a self-sustained equilibrium without external pressure at $T=0$. So the natural value for $\epsilon_{\rm vac}$ at $T=0$ in such a system in equilibrium is $\epsilon_{\rm vac}=0$. At $T\neq 0$, the thermal fluctuations, or quasi-particle excitations, lead to a matter pressure $p_M$, which is compensated by a non-zero vacuum pressure such that $p_{\rm vac}+p_M=0$. The vacuum energy therefore naturally evolves towards the value $\epsilon_{\rm vac}= p_M$ in equilibrium~\cite{Barcelo:2006cs}. The microscopic constituents of the system automatically adjust to obey the macroscopic thermodynamic rule, and there is no need to know the precise microscopic constitution of the system to calculate these macroscopic equilibrium quantities. The evolution towards this equilibrium, however, does depend on the microscopic constitution of the system. Still, a lot can be learned about the cosmological constant at a classical macroscopic level just by making some basic thermodynamic assumptions about the vacuum (e.g., by assuming that some generalized thermodynamic quantity must exist which is conserved because of the observed Lorentz invariance of the quantum vacuum, see~\cite{KlinkhamerVolovik2008,KlinkhamerVolovik2011a}). 

The cosmological constant mystery then becomes a lot less unsurmountable: From having to explain
why the cosmological constant is more than a hundred orders of magnitude smaller than its theoretically expected value, it is reduced to having to explain why it is slightly bigger than the equilibrium value which would exactly cancel the matter contribution: $\Omega_\Lambda\approx 0.7$ versus $\Omega_M\approx 0.3$. 
 So, the condensed matter approach offers at least a
qualitative framework to understand the problem of dark energy. 

Before explaining what this has to do with time, a few more remarks with respect to emergent spacetimes and their Lorentzian character might be useful.

\section{Emergent spacetimes and Lorentz invariance}
\label{S:lorentz_invariance}
From a relativistic point of view, the essence of physical time lies, first, (at the level of special relativity) in the fact that we live in a Lorentzian spacetime, i.e., a spacetime with a metric of signature $(- + + +)$. Second, general relativity adds to this the gravitational red- and blueshifts due to inhomogeneities in the gravitational field. These are caused by the matter/energy distribution, and become particularly acute near regions of dense distribution, e.g. near a black hole. 

We have seen in the previous section that the Lorentzian character of a spacetime can arise even when the underlying structure in itself is not Lorentzian, but can even be Newtonian, with an absolute time defined by the ``laboratory'' setting in which the atoms composing the microscopic condensed matter system live.\footnote{Mathematically, one could still define Lorentz transformations for such a system. However, the relativistic ``corrections'' compared to the Newtonian physics obtained from Galilean transformations would be irrelevant in practice. One may think, e.g., of a phase transition in a background system where all the velocities involved are {\it necessarily} much smaller than the relativistic speed characteristic of the background spacetime. This in fact is what happens in most laboratory systems which display effective acoustic gravity, and where $c_{\rm sound}\ll c_{\rm light}$. For all practical purposes, the background system may therefore be described as Newtonian, even though the ``internal'' physics in the effective gravity is naturally Lorentzian and governed by $c_{\rm sound}$. Note that it is not {\it required} for the emergence of an effective acoustic gravity that the background system be Newtonian. Analogue gravity also emerges, e.g., in relativistic Bose-Einstein condensates~\cite{Fagnocchi:2010sn}.} There are also other possibilities. Mathematically speaking, it is perhaps not so difficult to obtain an effective low-energy Lorentzian structure from a global ``timeless'' one. The following example illustrates this~\cite{barcelo-essay}.

A wave equation of the type
\begin{equation}
(-\partial_t^2+c^2\partial_x^2+c^2\partial_y^2+c^2\partial_z^2)\phi=0~,
\end{equation}
(or \eqref{alembert} in the more general case of a curved spacetime) is a hyperbolic partial differential equation. Changing the sign of the first term would lead to
\begin{equation}\label{parabolic}
(\partial_t^2+c^2\partial_x^2+c^2\partial_y^2+c^2\partial_z^2)\phi=0~,
\end{equation}
an elliptic partial differential equation. It is also often encountered in physics, for example in the Poisson equations of electrostatics, but has a totally different behaviour from the previous type. The mere change of sign of the first term implies that the temporal character of the coordinate $t$ is completely lost: in \eqref{parabolic}, $t$ behaves exactly in the same way as $x$, $y$ and $z$. So if the latter are spatial coordinates, then so is $t$ (modulo $c$ which is now merely a conversion factor between the unit of $t$ and the unit of the other coordinates, but cannot be interpreted as a velocity in any way). The same would be true if replacing the second-order partial derivatives $\partial_t^2,\partial_x^2,...$ by fourth-order ones: $\partial_t^4,\partial_x^4...$.

Now let us write down the following equation:
\begin{equation}
a(\partial_t^4+c^4\partial_x^4+c^4\partial_y^4+c^4\partial_z^4)\phi
+(-\partial_t^2+c^2\partial_x^2+c^2\partial_y^2+c^2\partial_z^2)\phi=0~,\label{fourth-order}
\end{equation}
where we assume that the (dimensionful) prefactor $a$ is given by
\begin{equation}
 a=(T/T_{\sf Planck})\tau^2_{\rm ch}
\end{equation}\label{prefactor-a}
with $\tau_{\rm ch}$ some characteristic $t$-scale. As such, Eq.~\eqref{fourth-order} is of the elliptic type just described, and so $t$ is a coordinate that behaves exactly as $x$, $y$ and $z$ do, and $c$ is just a dimensional constant without any possible interpretation of velocity.\footnote{In fact, at this point, $\tau_{\rm ch}$ can best be interpreted as a length scale, i.e. $\tau_{\rm ch}=\xi_{\rm ch}/c$ with $\xi_{\rm ch}$ some characteristic length scale of the system and $c$ a dimensional conversion factor.} 
However, at low temperatures, when $T$ becomes much lower than the Planck temperature $T_{\sf Planck}$, the prefactor $a$ becomes small, and the second part in~\eqref{fourth-order} can become dominant.\footnote{The fourth-order derivatives of the first part of Eq.~\eqref{fourth-order} imply that the global behaviour will in general be determined by it, and not by the second part. The conditions for the second part to become dominant in the limit when $a\to 0$ are actually mathematically quite subtle, but this is just meant as a simple pedagogical example to illustrate the point of obtaining a hyperbolic structure from an underlying non-hyperbolic one. More involved examples, including a discussion of the mathematical conditions for the obtention of a low-energy hyperbolic structure, can be found in \cite{barcelo-essay}.}
You then obtain a Lorentzian structure, and so it is perfectly legitimate to interpret $t$ as a time coordinate and $c$ as a velocity.

This was of course just a crude mathematical model. However, one often encounters claims that quantum gravity is or should be timeless, and that ``time'' is just a property that arises in the low-energy limit. This claim is in fact usually made in a sense quite different from the example just given\footnote{General Relativity can be formulated as a gauge theory, and should therefore be invariant under the transformations of the relevant gauge group, namely the diffeomorphism group. For our discussion, the relevant issue is that physical states which differ only by a time reparametrization should be physically equivalent. One can take this as a fundamental point when attempting to quantize GR, which leads to the idea that time should be absent altogether in a fundamental (``quantum'') description of gravity. The problem then is how to recover time at the classical, ``effective'' level, and in particular how the evolution of the universe comes about. See e.g. \cite{Kuchar:1991qf,Isham:1992ms} for broad reviews on the problem of time in quantum gravity, including a dicussion of timeless models, and \cite{Barbour:2009zd,Rovelli:2009ee} for introductions to two of the more popular approaches to timeless (quantum) gravity.}. Nevertheless, the above illustrates that obtaining a universe with a (relativistic) time conception (or more precisely: a metric with Lorentzian signature) from a timeless ``absolute'' law is perhaps not as hard as one might expect at first sight, at least in a mathematical sense. Also, since in the last example we could have replaced the first part of the equation by basically any equation preceded by $a$ and still obtain the same low-energy limit, it also illustrates that a variety of microscopic theories could lead to a low-energy effective Lorentzian spacetime in the adequate limit. The really hard part of quantum gravity, if the emergent point of view of condensed matter is relevant, is precisely that there does not exist any quantisation procedure, not even in principle, leading from the ``quasi-particle'' excitations that we experience in our low-energy physics (and the associated gauge fields) to the microscopic ``atoms'' of the quantum vacuum.

From the previous example, one might wonder how the transition between the low-energy Lorentzian effective spacetime and the very different high-energy or microscopic physics takes place. Does the relativistic spacetime suddenly disappear or is there a smooth transition? Condensed matter models again provide us with some useful clues. 

\section{Condensed matter models and Lorentz invariance}

In the past few years, intense experimental attention has been paid to the possibility that Lorentz invariance might be an effective low-energy phenomenon, broken at high energies~\cite{Jacobson:2005bg}. At the moment, no indication has been found that this should be the case, and actually there exist very stringent bounds on possible Lorentz violations at the Planck scale~\cite{Liberati:2013xla}. So maybe quantum gravity should include Lorentz invariance from first principles? In string theory and canonical quantum gravity, it is not really clear whether Lorentz violations are to be expected at high energies or not. In scenarios of emergent gravity
based on condensed matter analogies, the situation is much clearer: Lorentz invariance is a low-energy effective symmetry, and so
it is expected to break at some scale, although not necessarily
related to (and therefore possibly much higher than) the Planck
scale~\cite{Klinkhamer:2005cp}. The way in which the low-energy relativistic spacetime gradually makes place for the microstructure can be understood as follows.

Phenomenologically speaking, Lorentz breaking can be described simply by the following power law for the dispersion relation between the energy $E$ and the momentum $p$ (we consider massless particles and write $c$ for the invariant speed of the theory, be it a speed of light or a speed of sound):
\begin{eqnarray}
 E^2=c^2p^2+\alpha c^2p^4/p_{LV}^2 ~~(+~ \text{higher-order terms})
\end{eqnarray}
where the subscript $LV$ indicates the Lorentz violation scale, and $\alpha=\pm 1$ (we assume that uneven powers of $p$ are ruled out to lowest order, since they would lead to parity violation, i.e., a breaking of the symmetry under spatial reflection, at a much more fundamental level than the weak violations observed up to now). 
%
%

\subsection{Bose-Einstein condensates}
The Bogoliubov dispersion relation for Bose-Einstein condensates, in terms of the frequency $\omega$ and the wave number $k$, is
\begin{eqnarray}
\omega^2=c^2k^2+\frac{1}{4}c^2\xi^2k^4,
\end{eqnarray}
where $\xi\equiv \hbar/mc$ is the healing length of the condensate (roughly speaking, the distance needed for the condensate to smoothen out a sharp inhomogeneity in the atomic density). It is not obvious how to connect the Lorentz violation scale $k_{LV}=2/\xi$, or alternatively $E_{LV}=\hbar c/k_{LV}=2mc^2$, with the ``Planck scale'' of the theory. A naive reasoning could be the following. The Planck scale is the scale at which deviations from the classical picture become important. Then one might be tempted to identify the Lorentz violation scale with the Planck scale $k_P$, and so the stringent experimental bounds on Lorentz violation at the Planck scale would seemingly rule out an approach based on a BEC analogy. 

However, one should take care with this interpretation for two reasons. First, the BEC model is a model for the gravitational sector of the quantum vacuum only, and (as we already pointed out earlier) probably in the first place a toy model, so we cannot expect it to reproduce all features of the real quantum vacuum. In particular, the bosonic degrees of freedom included in the BEC model might be formed by effective coupling between fermionic degrees of freedom (through the formation of Cooper pairs, for example), or they might co-exist with other (fermionic) degrees of freedom. In both cases, information about the fermionic sector might be necessary to define and establish the hierarchy of the precise characteristic scales involved in the system. Second, even considering only the simple BEC model, already various characteristic scales can be constructed from the fundamental parameters of the microscopic theory: the Planck constant $\hbar$, the mass $m$ of the condensate atoms, their density $\rho$ (or the interatomic distance $a_0\sim \rho^{-1/3}$), and the interaction potential $U$ (for weakly interacting systems such as BECs in dilute gases, one has $U({\bf r})\approx U\delta({\bf r})$, with $U\propto a_s$, the $s$-wave scattering length; note that $a_s\ll a_0$ due to the weakness of the interaction). One can for example construct a second characteristic energy scale $E_{ch2}=\hbar c/a_0$, with $c=\sqrt{U\rho/m}$. $E_{ch2}$ can be interpreted as the energy scale at which the granularity of the vacuum becomes significant. In Bose gases, in general, $mca_0/\hbar\ll 1$, and hence $E_{LV}\ll E_{ch2}$, indicating that Lorentz violations are expected at much lower energies than the energy at which the discreteness of the vacuum becomes apparent. 

The main lesson to be drawn from this example is simply that naive dimensional estimates indicating that quantum gravity effects should be expected around ``{\it the} Planck scale'' $E_P=\sqrt{\hbar c^5/G}$, with $G$ the gravitational constant, are indeed naive. Different types of quantum gravity phenomenology might be characterised by different, mutually independent energy scales, which are not necessarily accessible to an internal observer who is limited to the effective low-energy physics. 

\subsection{Fermionic vacua}\label{SS:fermionic_vacua}
In a condensed-matter scenario where the fundamental degrees of freedom of the microscopic theory are fermionic, such as the Fermi-point scenario, the gravitational and gauge bosons are composite or collective excitations based on these fundamental fermions. In such a scenario, the ``Planck scale'' could be understood as the energy scale above which the bosonic content of the low-energy theory starts to dissolve into its fundamental fermionic components~\cite{Volovik:2003fe}, and---as stressed above---this $E_{Pl}$ can be quite different from the Lorentz-violation scale $E_{LV}$. When calculating the effective action for the bosonic fields, the result will depend on the hierarchy between $E_{LV}$ and $E_{Pl}$. In particular, when $E_{LV}\ll E_{Pl}$ (as is the case in $^3$He-A, and also in BECs when $E_{ch2}$ is interpreted as the Planck scale), then fermions with energies above the Lorentz violation scale contaminate this effective action with non-covariant (``hydrodynamic'') terms, and the result will be an action which strongly violates diffeomorphism invariance (see \cite{Volovik:1986ix} for the explicit case of  $^3$He-A, ~\cite{Barcelo:2010vc} for general considerations on the mechanism and the role of the different scales involved, and~\cite{Barcelo:2007iu} for the difference between diffeomorphism at the kinematical level---which is obeyed---and diffeomorphism invariance at the dynamical level---where the challenge lies). Therefore, to obtain Einstein gravity, the ultraviolet cut-off scale for the fermions must be (much) lower than the Lorentz violation scale. So a good condensed-matter-like model for emergent gravity would require a system in which these characteristic scales are reversed with respect to the case of $^3$He-A (and BECs): $E_{Pl}\ll E_{LV}$, in agreement with astrophysical and cosmological observations on (the absence of) Lorentz violations.

In fact, the previous exercise of assuming that the gauge bosons are collective excitations from fundamental fermions can be done explicitly with respect to the interactions of the standard model~\cite{Klinkhamer:2005cp}, leading to the possibility of obtaining ``GUT''-like unification (in the sense of an approximate merging of the running coupling constants) without the need to impose additional symmetries. Comparison of the parameters of the model with the experimental knowledge of these coupling constants also allows to estimate the relation between the bosonisation scale for these interactions and the Lorentz violation scale $E_{LV}$ in such a model.\footnote{The bosonisation scale for the standard model interactions considered in~\cite{Klinkhamer:2005cp} need not coincide with the gravitational bosonisation scale. In fact, \cite{Klinkhamer:2005cp} finds \mbox{$\sim 10^{13}-10^{15}$GeV} for the former, i.e. $10^{-6}-10^{-4} E_{Pl}$. Note that, in a laboratory condensed matter system such as $^3$He-A, different types of collective bosons also need not necessarily appear at the same temperature, external magnetic field etc.} Assuming that our gravitational bosonisation scale corresponds to the usual estimate for the Planck scale (\mbox{$\sim 10^{19}$ GeV}), a conservative estimate gives $E_{Pl}/E_{LV}\lesssim 10^{-8}$ (the exact proportion seems to depend on the number of fermion families $N_F$ --- for $N_F=3$, $E_{Pl}/E_{LV}\sim 10^{-25}$ is obtained), again in qualitative agreement with current boundaries on Lorentz violations.

In any case, the additional lesson with respect to the case of BECs might well be that, in a sufficiently complex system, the various characteristic scales could in a certain sense ``conspire'' to protect the effective low-energy symmetries such as Lorentz invariance, and hide the microscopic physics from a low-energy observer. Indeed, if the compositeness scale of the bosons provides a cut-off for low-energy beings such as ourselves, while the Lorentz violation scale lies at much higher energies, then the latter is suppressed from observation by at least the huge factor $E_{LV}/E_{Pl}$. This would then mean that relativistic spacetimes subsist well above the Planck scale, even though the ``quantum gravity'' theory itself (i.e., the theory describing the microscopic constituents of the spacetime condensate) does not obey Lorentz invariance. What would physics between the Planck and the Lorentz violation scale look like, then? An intriguing possibility is that the effective spacetime itself would survive unaltered, but that gravity would be modified at energies $E_{Pl}< E< E_{LV}$ and might even vanish completely before $E_{LV}$ is reached. One would then be left with a relativistic spacetime in the pure sense of special relativity~\cite{Volovik:2003nx}: a non-gravitating, but still well-defined Lorentzian spacetime, where the invariant speed $c$, which at low energy was the signalling velocity of the (massless) bosons, is now the limiting velocity of the fermions. One might think that this would violate all we know about relativistic spacetimes in strong gravitational fields such as  black holes. However, a careful analysis shows that this would only modify the physics inside the black hole but is otherwise in perfect agreement with all the observed physics, and could in fact even open up new scenarios for solving some of the paradoxes associated with black-hole physics, see~\cite{Barcelo:2010vc}.

Curiously, the lesson from condensed matter seems to show that the order of fundamentality of our best theories 
of spacetime might have to be revised, or at the very least should be taken with caution. It is usually held that Newtonian space and time is just a low-velocity/low-energy limiting case of the more fundamental flat Minkowski spacetime, which in its turn is a low-energy limiting case of the more fundamental general relativistic curved spacetimes. The scenario just mentioned would imply that general relativistic curved spacetimes would, at energies above the Planck scale, give way to a flat Minkowski spacetime. At even higher energies, above the Lorentz violation scale, the real fundamental theory of the microconstituent atoms of spacetime might either be timeless, in the sense described in the previous section, but it could also be simply Newtonian, with an absolute time defined by a clock on the wall of ``God's laboratory'', as in a real condensed matter system in a real world laboratory.

\section{Challenges for emergent gravity}
\label{S:dynamics}
It is only fair to say that every current approach to quantum or emergent gravity has its pros and cons, and its important open questions. In the case of condensed-matter-inspired emergent gravity, the first key challenge lies in obtaining the desired dynamical equations. If one considers that gravity (and perhaps also the other fundamental gauge interactions) emerges through a kind of condensation process from a fundamental fermionic system, then---as mentioned in the previous section---one can in principle calculate the effective action for the boson fields {\it \`a la} Sakharov~\cite{Sakharov:1967pk} (or {\it \`a la} Zel'dovich for electromagnetism~\cite{zeldovich} -- see also~\cite{Barcelo:2014yna}). The central idea in Sakharov's induced gravity is that the quantum fluctuations of a classical curved Lorentzian spacetime automatically lead to an action containing terms that correspond to the cosmological constant and to the Einstein-Hilbert action, and hence to the Einstein equations, plus higher-order corrections. However, an essential assumption is that the spacetime has no prior dynamics~\cite{Visser:2002ew}. In the condensed matter analogies, on the contrary, the vacuum is governed by prior dynamics, e.g. the Euler and continuity equation of hydrodynamics. These cannot just be ignored since they are precisely crucial for the emergence of the effective metric. This is one of the reasons why implementing Sakharov's idea is far from trivial in such a setting.

Another way to see this is the following. Because of the argument discussed in Sec.~\ref{SS:fermionic_vacua}, when $E_{LV}\ll E_{Pl}$, the Sakharov-style effective action will contain strongly non-diffeomorphism-invariant terms. However, in all known laboratory condensed matter systems, the emergence of Lorentz invariance is precisely a phenomenon which occurs in the low-energy limit within the (already) condensed phase. If the cut-off scale $E_{Pl}$ corresponds to the condensation scale, then it seems impossible to obtain $E_{LV}\gg E_{Pl}$ through a laboratory-style condensation mechanism. To solve this problem, perhaps one should look for a background system that is already Lorentz invariant from the start~\cite{Belenchia:2014hga}. But this is clearly not in line with the spirit of emergence that has been presented here. Alternatively, one could look for a different topological mechanism, e.g. related to Fermi-point splitting and merging~\cite{Volovik:2013cga}. Or perhaps one should imagine a multi-step condensation process, in which (from high to low energy) the first condensation produces the effective Lorentz symmetry (but, at the dynamical level, retains a memory of the background), while the resulting collective excitations again condense to produce a diffeomorphism invariant action. If the energy separation between both processes is sufficiently large, or some screening mechanism arises inbetween, then perhaps this could lead to a decoupling of the final effective theory from the original background such that the diffeomorphism-breaking terms would vanish, or at least be small enough as to be in agreement with observations. 

But this is so far an open challenge, and in any case would not be the end of the story. Gravity being a long-range force, the corresponding gauge boson---i.e., the graviton---should in principle be massless. Moreover, we know from the equivalence principle that gravity couples not just to rest mass, but to energy (i.e., to the stress-energy tensor), and this is one of the most well-tested principles in physics. This implies that gravity should be mediated by a spin-2 field (see e.g. the introduction to~\cite{Feynman:1996kb}), which (as just stated) should also be massless. Now, according to the Weinberg-Witten theorem~\cite{Weinberg:1980kq}, in a perfectly Lorentz invariant setting, such a massless spin-2 field (fundamental or composite) must necessarily be gauge invariant (in particular: diffeomorphism invariant), with no spin 0 or spin 1 components. In our case, one could dismiss the Weinberg-Witten theorem since Lorentz invariance is broken at high energy~\cite{Jenkins:2009un}. But this implies (according to the argument of the previous paragraph) that diffeomorphism invariance will also be an emergent, non-exact symmetry. However, even an arbitrarily small breaking of diffeomorphism invariance leads to the presence of massive spin-0 and/or spin-1 graviton components, contrarily to what we just stated as fundamental characteristics of gravity. Even if these massive spin-0 and/or spin-1 graviton components were extremely small, they could still be rather problematic, as they produce so-called Boulware-Deser ghosts (negative energy states)~\cite{Boulware:1973my}, which in general do not disappear even in the $m\to 0$ limit. Recent work has shown that these ghosts might be absent in particular settings, leading to a ghost-free ``de Rham-Gabadadze-Tolley'' massive gravity theory~\cite{deRham:2010kj}. Whether the conditions for such massive gravity are realistic is a subject of active debate. But, interestingly, such ghost-free massive gravity theories are intimately related to bi-metric gravity theories of which the effective spacetimes emerging in condensed-matter analogies are a (simplified) example~\cite{Baccetti:2012bk}. 

A basic illustration of some of the aspects just mentioned is again to be found in $^3$He-A. The analogue of gravitons in $^3$He-A (perturbations of the effective quasiparticle metric) correspond to so-called clapping modes~\cite{volovik1987a,volovik1987b}, which have been experimentally detected and studied as of the 1980s~(\cite{halperin-book} and references therein). However, the analogy is imperfect, as there are indeed two collective spin-2 modes in $^3$He-A, but these are massive, and there is also a massive spin-0 component.\footnote{These analogue graviton masses are curiously related to the value of the analogue cosmological constant in $^3$He-A, see~\cite{Jannes:2011em}} Given the previous discussion, this should not come as a surprise, since $^3$He-A is obviously not a diffeomorphism invariant structure.

The relevance on the discussion of emergent time is that, if diffeomorphism invariance turns out {\it necessarily} to be an exact symmetry, as the above problems perhaps suggest, then this implies that the fundamental theory must also necessarily be invariant (at all scales) under (local) reparametrizations of the time coordinate, and this would impose strong limits on the character of the fundamental theory. And indeed, the whole previous discussion might sound like an unsurmountable barrier for emergent gravity as has been described so far. But it is also worth remembering that perfect symmetries do not seem to be very popular with Mother Nature (CPT-invariance being, so far, perhaps the only reasonably well-established candidate). So, rather than assuming one (diffeomorphism invariance) or even two exact symmetries (Lorentz and diffeomorphism invariance) in a single shot in order to avoid the problems just mentioned, it might be worth studying whether all these problems are truly a no-go, or rather a ``technical challenge'' towards a more complete emergent theory of gravity in which both Lorentz and diffeomorphism invariance are emergent symmetries.

For more extensive analyses of the challenges associated with emergent gravity, see e.g.~\cite{Sindoni:2011ej,Carlip:2012wa}.

\section{Some final comments}
\label{S:conclusion}
Is the spacetime of our universe really composed of a condensed matter system? It might be completely wrong to think of spacetime as composed of material atoms, more or less localised in their own Newtonian (or other) inertial frame, even if there is currently no experiment that excludes such a possibility, since this ``absolute'' high-energy inertial frame would be totally unreachable at the low energies at which we have probed the laws of nature up to now. But it might nevertheless be very instructive to consider the possibility that, whatever the constitution of the fundamental degrees of freedom of the quantum vacuum composing the universe, these degrees of freedom are nearly completely frozen out at the extremely low temperatures (compared to their characteristic scale) that are present in most of the actual universe. The physics at such extremely low temperatures might then well be governed by collective excitations and their emergent symmetries in a way very similar to what occurs in real laboratory condensed matter systems.

I have therefore briefly described emergent scenarios for gravity based on condensed matter models, and focused on two particular cases: Bose-Einstein condensates and $^3$He-A. A key point is that, in these condensed matter scenarios, Lorentz invariance is an effective, low-energy symmetry which is expected to break at high energy. Crucially, in such scenarios for emergent gravity, there is no reason to expect a single characteristic energy scale for quantum gravity (``{\it the} Planck scale''), so various aspects of quantum gravity phenomenology could be associated to different energy scales. In particular, the energy scale of Lorentz violation is expected to be many orders of magnitude higher than the cut-off scale for the effective low-energy physics, which in a theory with a fermionic vacuum could for example be the bosonic compositeness scale. Then, the effective low-energy symmetries would be protected by the proportion between these two energy scales. This would mean that any modification of the continuous effective Lorentzian spacetime at high energies would be extremely hard to detect experimentally, much harder than in the usual scenarios based on a single characteristic Planck scale for all quantum gravitational effects. This should of course not be taken as a defeatist attitude, but on the contrary as an additional stimulation to further develop ingenuous experiments at high energies, and see which properties of the microstructure of spacetime we can infer from their results. The bottom-line might then seem a bit disappointing: we are still very far away from understanding the true origin of physical time, and many experimental advances will be needed before we start doing so. But then again, we have only approximately understood less than 5\% of the total energy content of our observable universe. So maybe thinking that we could already begin to understand the true essence of time was slightly over-optimistic in any case.

\begin{acknowledgements}
I thank F. Barbero, C. Barcel\'o and G.E. Volovik for useful comments. Financial support was provided by the Spanish MICINN through the project FIS2011-30145-C03-01.
\end{acknowledgements}



\begin{thebibliography}{99}
\bibitem{Volovik:2003fe}
  G.~E.~Volovik,
  {\it The Universe in a helium droplet,}
  Clarendon Press, Oxford (2003).
\bibitem{Unruh:1980cg} 
  W.~G.~Unruh,
  ``Experimental black hole evaporation?''
  Phys.\ Rev.\ Lett.\  {\bf 46}, 1351 (1981).
\bibitem{Visser:1997ux}
  M.~Visser,
  ``Acoustic black holes: Horizons, ergospheres, and Hawking radiation,''
  Class.\ Quant.\ Grav.\  {\bf 15}, 1767 (1998).
\bibitem{Barcelo:2005fc}
  C.~Barcel\'{o}, S.~Liberati and M.~Visser,
  ``Analogue gravity,''
Living Rev.\ Rel.\  {\bf 14}, 3 (2011).
\bibitem{Lahav:2009wx} 
  O.~Lahav, A.~Itah, A.~Blumkin, C.~Gordon and J.~Steinhauer,
  ``Realization of a sonic black hole analogue in a Bose-Einstein condensate,''
  Phys.\ Rev.\ Lett.\  {\bf 105}, 240401 (2010)
\bibitem{Steinhauer:2014dra} 
  J.~Steinhauer,
  ``Observation of self-amplifying Hawking radiation in an analog black hole laser,''
  Nature Phys.\  {\bf 10}, 864 (2014)
\bibitem{Sakharov:1967pk}
  A.~D.~Sakharov,
  ``Vacuum quantum fluctuations in curved space and the theory of gravitation,''
  Sov.\ Phys.\ Dokl.\  {\bf 12}, 1040 (1968)
  [Dokl.\ Akad.\ Nauk Ser.\ Fiz.\  {\bf 177}, 70 (1967)].
\bibitem{Girelli:2008gc}
  F.~Girelli, S.~Liberati and L.~Sindoni,
  ``Gravitational dynamics in Bose Einstein condensates,''
  Phys.\ Rev.\  D {\bf 78}, 084013 (2008).
\bibitem{Girelli:2008qp} 
  F.~Girelli, S.~Liberati and L.~Sindoni,
  ``Emergence of Lorentzian signature and scalar gravity,''
  Phys.\ Rev.\ D {\bf 79}, 044019 (2009)
\bibitem{Barcelo:2014yna} 
  C.~Barcel\'o, R.~Carballo-Rubio, L.~J.~Garay and G.~Jannes,
  ``Electromagnetism as an emergent phenomenon: a step-by-step guide,''
  New J. of Phys. (in press).
\bibitem{Padmanabhan:2008if} 
  T.~Padmanabhan,
  ``Dark Energy and its Implications for Gravity,''
  Adv.\ Sci.\ Lett.\  {\bf 2}, 174 (2009)
\bibitem{Weinberg:1988cp}
  S.~Weinberg,
  ``The cosmological constant problem,''
  Rev.\ Mod.\ Phys.\  {\bf 61}, 1 (1989).
\bibitem{Volovik:2006bh}
  G.~E.~Volovik,
  ``Vacuum energy: Myths and reality,''
  Int.\ J.\ Mod.\ Phys.\  D {\bf 15}, 1987 (2006).
\bibitem{Jannes:2011em} 
  G.~Jannes and G.~E.~Volovik,
  ``The cosmological constant: A lesson from the effective gravity of topological Weyl media,''
  JETP Lett.\  {\bf 96}, 215 (2012)
\bibitem{Barcelo:2006cs}
  C.~Barcel\'{o},
  ``Cosmology as a search for overall equilibrium,''
  JETP Lett.\  {\bf 84}, 635 (2007).
\bibitem{KlinkhamerVolovik2008}
 F.R. Klinkhamer and G.E. Volovik,  
 ``Self-tuning vacuum variable and cosmological constant,'' 
  Phys. Rev. D {\bf 77}, 085015 (2008).
\bibitem{KlinkhamerVolovik2011a} 
  F.R. Klinkhamer and G.E.Volovik,
``Dynamics of the quantum vacuum: Cosmology as relaxation to the equilibrium
state,''
 J. Phys. Conf. Ser. {\bf 314}, 012004 (2011).
\bibitem{Fagnocchi:2010sn} 
  S.~Fagnocchi, S.~Finazzi, S.~Liberati, M.~Kormos and A.~Trombettoni,
  ``Relativistic Bose-Einstein Condensates: a New System for Analogue Models of Gravity,''
  New J.\ Phys.\  {\bf 12}, 095012 (2010)
\bibitem{barcelo-essay}
  C.~Barcel\'{o},
``Lorentzian Space-Times from Parabolic and Elliptic Systems of PDEs,''
 in: V.~Petkov (ed.), {\it Relativity and the Dimensionality of the World}, Springer (2008).
\bibitem{Kuchar:1991qf}
  K.~V.~Kuchar,
  ``Time and interpretations of quantum gravity,''
  Int.\ J.\ Mod.\ Phys.\ Proc.\ Suppl.\ D {\bf 20}, 3 (2011).
\bibitem{Isham:1992ms} 
  C.~J.~Isham,
  ``Canonical quantum gravity and the problem of time,'' arXiv:gr-qc/9210011.
\bibitem{Barbour:2009zd} 
  J.~Barbour,
  ``The Nature of Time,''
  arXiv:0903.3489 [gr-qc].
\bibitem{Rovelli:2009ee}
 C.~Rovelli,
  ``'Forget time',''
  Found.\ Phys.\  {\bf 41}, 1475 (2011)
\bibitem{Jacobson:2005bg}
  T.~Jacobson, S.~Liberati and D.~Mattingly,
  ``Lorentz violation at high energy: Concepts, phenomena and astrophysical
  constraints,''
  Annals Phys.\  {\bf 321}, 150 (2006).
\bibitem{Liberati:2013xla} 
  S.~Liberati,
  ``Tests of Lorentz invariance: a 2013 update,''
  Class.\ Quant.\ Grav.\  {\bf 30}, 133001 (2013)
\bibitem{Klinkhamer:2005cp}
  F.~R.~Klinkhamer and G.~E.~Volovik,
  ``Merging gauge coupling constants without grand unification,''
  Pisma Zh.\ Eksp.\ Teor.\ Fiz.\  {\bf 81}, 683 (2005)
  [JETP Lett.\  {\bf 81}, 551 (2005)].
\bibitem{Volovik:1986ix} 
  G.~E.~Volovik,
  ``Chiral Anomaly and the Law of Conservation of Momentum in $^{3}$He-A,''
  JETP Lett.\  {\bf 43}, 551 (1986)
  [Pisma Zh.\ Eksp.\ Teor.\ Fiz.\  {\bf 43}, 428 (1986)].
\bibitem{Barcelo:2010vc} 
  C.~Barcel\'o, L.~J.~Garay and G.~Jannes,
  ``Quantum Non-Gravity and Stellar Collapse,''
  Found.\ Phys.\  {\bf 41}, 1532 (2011)
\bibitem{Barcelo:2007iu} 
  C.~Barcel\'o and G.~Jannes,
  ``A Real Lorentz-FitzGerald contraction,''
  Found.\ Phys.\  {\bf 38}, 191 (2008)
\bibitem{Volovik:2003nx}
  G.~E.~Volovik,
  ``Vacuum energy and universe in special relativity,''
  JETP Lett.\  {\bf 77}, 639 (2003)
  [Pisma Zh.\ Eksp.\ Teor.\ Fiz.\  {\bf 77}, 769 (2003)].
\bibitem{zeldovich}
Y.~B.~Zel'dovich, 
``Interpretation of Electrodynamics as a Consequence of Quantum Theory,''
 JETP Lett. {\bf 6}, 345 (1967).
\bibitem{Visser:2002ew} 
  M.~Visser,
  ``Sakharov's induced gravity: A Modern perspective,''
  Mod.\ Phys.\ Lett.\ A {\bf 17}, 977 (2002)
\bibitem{Belenchia:2014hga} 
  A.~Belenchia, S.~Liberati and A.~Mohd,
  ``Emergent gravitational dynamics in a relativistic Bose-Einstein condensate,''
  Phys.\ Rev.\ D {\bf 90}, 104015 (2014)
\bibitem{Volovik:2013cga} 
 G.~E.~Volovik,
 ``The Topology of the Quantum Vacuum,''
 Lect.\ Notes Phys.\  {\bf 870}, 343 (2013).
\bibitem{Feynman:1996kb} 
  R.~P.~Feynman, F.~B.~Morinigo, W.~G.~Wagner and B.~Hatfield,
  {\it Feynman lectures on gravitation,} Addison-Wesley (1995) 
\bibitem{Weinberg:1980kq} 
 S.~Weinberg and E.~Witten,
``Limits on Massless Particles,''
 Phys.\ Lett.\ B {\bf 96}, 59 (1980).
\bibitem{Jenkins:2009un} 
  A.~Jenkins,
  ``Constraints on emergent gravity,''
  Int.\ J.\ Mod.\ Phys.\ D {\bf 18}, 2249 (2009)
 \bibitem{Boulware:1973my} 
  D.~G.~Boulware and S.~Deser,
  ``Can gravitation have a finite range?,''
  Phys.\ Rev.\ D {\bf 6}, 3368 (1972).
 \bibitem{deRham:2010kj} 
  C.~de Rham, G.~Gabadadze and A.~J.~Tolley,
  ``Resummation of Massive Gravity,''
  Phys.\ Rev.\ Lett.\  {\bf 106}, 231101 (2011)
  \bibitem{Baccetti:2012bk} 
  V.~Baccetti, P.~Martin-Moruno and M.~Visser,
  ``Massive gravity from bimetric gravity,''
  Class.\ Quant.\ Grav.\  {\bf 30}, 015004 (2013)
  \bibitem{volovik1987a}
G.~E.~Volovik,
``Singular behavior of the superfluid $^3$He-A at T=0 and quantum field theory,''
J. Low Temp. Phys. {\bf 67}, 301 (1987).  
  \bibitem{volovik1987b}
G.~E.~Volovik,
''Peculiarities in the dynamics of superfluid $^3$He-A: analog of chiral anomaly and of zero-charge,``
Sov. Phys. JETP {\bf 65}, 1193 (1987).
\bibitem{halperin-book}
W.P. Halperin, and L.P. Pitaevskii (eds.), {\it Helium three,} Elsevier (1990)
\bibitem{Sindoni:2011ej} 
  L.~Sindoni,
  ``Emergent Models for Gravity: an Overview of Microscopic Models,''
  SIGMA {\bf 8}, 027 (2012).
\bibitem{Carlip:2012wa} 
  S.~Carlip,
  ``Challenges for Emergent Gravity,''
  Stud.\ Hist.\ Philos.\ Mod.\ Phys.\  {\bf 46}, 200 (2014).


\end{thebibliography}
\end{document}